\documentclass[manuscript]{aastex}

\shorttitle{Carbon-rich RR Lyr type stars.}
\shortauthors{Wallerstein et al.}

\begin{document}


\title{Carbon-rich RR Lyr type stars.}


\author{George Wallerstein \altaffilmark{1}}
\affil{ Department of Astronomy, University of Washington, Seattle,
 WA 98195, wall@astro.washington.edu}

\author{V. V. Kovtyukh\altaffilmark{2} and S. M. Andrievsky\altaffilmark{2}}
\affil{Department of Astronomy and Astronomical Observatory,
Odessa National University, T.G. Shevchenko Park, 65014,
Odessa, Ukraine}
\email{val@deneb1.odessa.ua, scan@deneb1.odessa.ua}

\begin{abstract}
We have derived CNO abundances in 12 RR Lyrae stars. Four stars show [C/Fe]
near 0.0 and two stars show [C/Fe] = 0.52 and 0.65. Red giant branch stars,
which are known to be the predecessors of RR Lyrae stars, generally show a
deficiency of carbon due to proton captures during their evolution
from the main sequence up the giant branch. We suggest that the enhancement
of carbon is due to production during the helium flash combined with mixing
to the surface by vigorous convection induced by the flash itself.
\end{abstract}


\section{Introduction}
It is generally agreed that the RR Lyrae stars have evolved from the red
giant branch (RGB) to their present position on the horizontal branch (HB)
of globular clusters when the helium flash initiates the triple-alpha
reaction in the degenerate core. Because of the degeneracy, combined with
the great temperature dependence of the triple-alpha reaction, the core of
the red giant heats extremely rapidly leading to very vigorous convection
that must be treated by hydrodynamic methods rather than the conventional
theory of steady convection. Recent calculations have been described by
Dearborn, Lattanzio \& Eggleton (2006) as well as by Mocak et al. (2008).
There remain many uncertainties in the calculations but it is generally
agreed that the star is not disrupted. This is demonstrated by the very
existence of HB stars. At the same time the degree of mixing remains
uncertain with the distinct possibility that freshly minted $^{12}$C may
reach the surface. In addition the $^{12}$C must pass through a region
rich in hydrogen so it may be reprocessed to $^{13}$C and $^{14}$N.

As part of a small survey of the chemical composition of specially
selected RR Lyrae stars and comparison stars of known  metallicity we
have found at least two carbon-rich objects.

\section{Observations}

The spectra of our program stars were obtained with the echelle spectrograph
of the Apache Point Observatory (APO). By using a prism as cross-disperser
the APO echelle covers all wavelengths from 3500 \AA~ to 10400 \AA. However
the red-sensitive $2048 \times 2048$ chip has decreasing sensitivity below
4000 \AA~ and beyond 9000 \AA. The resolving power is about 35000. Exposure
times were usually about 20--30 minutes. We estimated the S/N ratio at the
continuum level depending upon the wavelength interval to be about 70--150
per pixel after combining multiple exposures that had been taken sequentially.
Table 1 contains the dates, JD, phases, and derived values of
T$_{\rm eff}$ and $\log~g$ for each phase.

\begin{table*}
\scriptsize
\caption[]{Observations of stars and their fundamental characteristics}
\begin{tabular}{ccccccccc}
\hline
Star    &Period (days)&Date      &  JD     &$\phi$   &T$_{\rm eff}$&$\log~g$&V$_{\rm t}$&Remarks\\
        &       &          &2450000+ &         &  K          &        &km s$^{-1}$&       \\
\hline
RS Boo  &0.377   & 2002-05-23&  2417.628& .440(b) & 6500 & 2.9 & 3.3 &                \\
        &     & 2006-03-21&  3816.745& .311(b) & 6700 & 3.5 & 3.5 &                \\
\hline
SW And  &0.442   & 2005-01-15&  3386.542& .797(c) & 6200 & 2.5 & 3.8 &            \\
\hline
AN Ser  &0.522   & 2002-05-23&  2417.664& .886(f) & 7500 & 2.8 & 4.8 &            \\
        &     & 2002-06-22&  2447.889& .780(f) & 6150 & 2.4 & 3.4 &            \\
        &     & 2002-06-23&  2448.638& .214(f) & 6300 & 2.6 & 3.7 &            \\
\hline
RR Lyr  &0.567   & 2003-06-14&  2805.790& .268(e) & 6200 & 2.6 & 3.8 &            \\
        &     & 2005-07-05&  3556.950& .448(e) & 6100 & 2.8 & 2.3 &            \\
\hline
UY Boo  &0.651   & 2002-05-23&  2417.733&         & 6200 & 2.0 & 2.8 &Na D1,D2 double\\
\hline
SU Dra  &0.660& 2006-03-21&  3816.731& .676(c) & 6100 & 2.5 & 3.5 &H$\alpha$ emission or double\\
\hline
VY Ser  &0.714   & 2002-03-02&  2336.876& .000(e) & 6800 & 2.0 & 3.0 &H$\alpha$ emission or double\\
        &     & 2002-05-22&  2417.688& .168(e) & 6350 & 2.0 & 3.0 &            \\
        &     & 2002-06-21&  2447.637& .107(e) & 6450 & 2.2 & 3.0 &            \\
        &     & 2002-06-22&  2448.689& .581(e) & 6000 & 2.3 & 3.7 &            \\
\hline
OX Her  &0.757   & 2005-06-18&  3539.757& .756(i) & 6000 & 2.7 & 3.3 &      \\
        &     & 2005-06-20&  3541.702& .324(i) & 6800 & 3.0 & 3.8 &      \\
\hline
V486 Her&0.806   & 2004-06-05&  3162.756& .260(d) & 6700 & 3.4 & 3.8 &  H$\alpha$ emission or double\\
        &     & 2005-03-26&  3456.880& .210(d) & 6700 & 3.3 & 3.5 &  H$\alpha$ emission or double\\
        &     & 2005-06-17&  3539.699& .972(d) & 7000 & 3.8 & 4.5 &  H$\alpha$ blue-shifted emission\\
        &     & 2005-06-19&  3541.762& .532(d) & 6200 & 3.2 & 4.0 &                         \\
\hline
XZ Cet  &0.823(a)& 2003-01-24&  2664.596& .167(h) & 6400 & 2.3 & 3.0 &                \\
        &     & 2003-02-07&  2678.562& .134(h) & 6450 & 2.2 & 2.8 &                \\
\hline
KP Cyg  &0.856   & 2005-11-13&  3687.606& .135(d) & 7050 & 2.6 & 2.5 &             \\
        &     & 2006-11-05&  4044.672& .299(d) & 6600 & 3.0 & 3.3 &             \\
        &     & 2007-05-01&  4221.963& .430(d) & 6450 & 3.4 & 3.5 &H$\alpha$ emission\\
        &     & 2005-09-24&  3637.607& .720(d) & 6300 & 2.4 & 3.2 &H$\alpha$ emission\\
        &     & 2006-10-03&  4011.720& .800(d) & 6650 & 3.0 & 4.2 &H$\alpha$ emission\\
        &     & 2005-09-13&  3626.711& .991(d) & 7400 & 3.0 & 2.5 &             \\
\hline
UY CrB  &0.929   & 2004-06-05&  3162.681& .491(g) & 6150 & 2.3 & 3.8 &             \\
        &     & 2005-03-26&  3456.989& .245(g) & 6700 & 2.4 & 3.4 &             \\
        &     & 2005-09-13&  3626.644& .839(g) & 6300 & 2.9 & 3.0 &             \\
\hline
\end{tabular}
\\
Remark: Phases are from:

a - XZ Cet was classified as an anomalous cepheid according to Teays \& Simon (1985), but that classification is uncertain.
b - Jones R.V. et al., 1987, ApJ 314, 605 
c - Liu T.\& Janes K.A., 1990, ApJ 354, 273
d - Loomis Ch., Schmidt E.G., Simon N.R., 1988, MNRAS 235, 1059
e - Takeda Y., Honda S., Aoki W., Takada-Hidai M., Zhao G., Chen Yu-Qin, Shi Jian-Rong, 2006, PASJ 58, 389
f - Lub J., 1977, A\&AS 29, 345 
g - Schmidt E.G., 2002, AJ 123, 965 
h - Schmidt E.G., 2002, AJ 123, 965 
i – GCVS, 1984, 1985

\end{table*}

\section{Data Reduction and Analysis}

\subsection{Data Reduction}

The program spectra were extracted from the raw frames using standard IRAF
procedures. The continuum level placement, wavelength calibration and
equivalent width measurements were performed with DECH20 code (Galazutdinov
1992).

\subsection{Atmospheric Parameters and Abundance Analysis}

The elemental abundances were derived using the Kurucz's WIDTH9 code with
atmosphere models interpolated from the ATLAS9 model grid. We used $\log~gf$
values derived from an inverted solar analysis (Kovtyukh \& Andrievsky 1999).

Atmosphere parameters (T$_{\rm eff}$, $\log~g$(sp), V$_{\rm t}$) were derived
by enforcing traditional spectroscopic criteria. Lines of Fe~I were forced
to yield zero slope in the relations between total iron abundance and
excitation potential. The total abundances of iron as predicted from Fe~I and
Fe~II lines were equalized by adjusting the model gravity (i.e. to yield
a "spectroscopic gravity").

\section{CNO abundances}

To derive the carbon abundance the following set of the lines was used:
6001.12 \AA, 6010.68 \AA, 6014.83 \AA, 6587.61 \AA, 6655.51 \AA, 7085.47 \AA,
7087.83 \AA, 7111.48 \AA, 7113.18 \AA, 7115.19 \AA, 7116.99 \AA, 8335.15 \AA.
The Nitrogen abundance was found from the lines 7442.29 \AA, 7468.30 \AA,
8216.34 \AA, 8242.39 \AA, 8629.16\AA, 8683.39, \AA, 8703.25 \AA,
8711.67 \AA, 8718.76 \AA. These lines are seen only in the stars with
a quite high metallicity. For the metal-poor ones only some of them were
present. The abundance of oxygen in the program stars was derived from the following lines: 6156.77 \AA, 6158.18 \AA, and
6300.30 \AA. 
Equivalent widths will be presented in a full paper on abundances of all
available elements in the stars in Table 1.

As an example, in Fig. 1 we show spectra in the region of the C~I lines near
7110 \AA. Note that all stars showing these lines were observed near the same
phase, 0.5, and hence have very nearly the same temperatures.

\begin{figure}
\resizebox{\hsize}{!}{\includegraphics{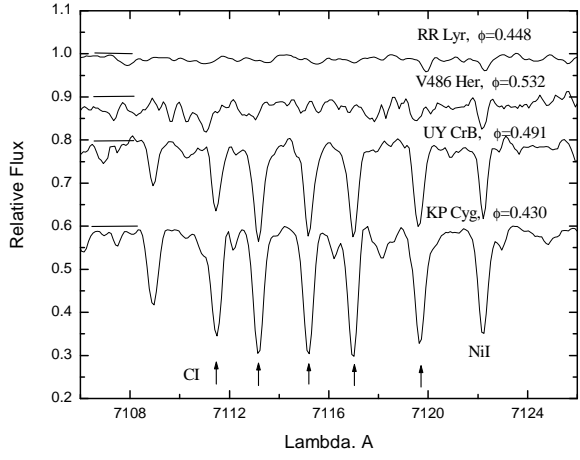}}
\caption[]{C~I lines in the program stars. Note that all stars were observed at nearly the same phase so their effective temperatures are similar.}
\end{figure}

The results for the CNO abundances in our program stars are given in Table 2
(we have shown the derived abundances on the usual scale letting logN(H)=12.0).

Only a few comparisons may be made with other analyses of RR Lyr star abundances.
For SW And Clementini et al. (1995) found [C/Fe] = -0.10 while we found $+$0.09.
The difference is almost entirely due to their solar abundance of C of 8.56 
(Anders and Grevesse 1990) and our more recent value of 8.39 (Asplund et al. 2005).
For [O/Fe] they found $-$0.03 and we found $+$0.23 which corresponds closely
to the revision of the solar oxygen abundance from 8.93 to 8.66.

It is also worth noting the carbon and s-process enhancements found by Preston et al.
(2006) in TY Gru, due probably to mass transfer from a companion. We have analysed
the heavy element content of the two C-rich stars, KP Cyg and UY CrB (to be reported
on later) and found no significant enhancements. 

\begin{table}
\scriptsize
\caption[]{CNO abundances}
\begin{tabular}{rrrrrrrr}
\hline
   Star  & $\log$N(C) & [C/Fe] &$\log$N(N) &
   [N/Fe] & $\log$N(O) & [O/Fe] & [Fe/H]\\
\hline
  RS Boo &8.34   &--0.06    &8.00   & +0.21&8.44&--0.21& +0.01 \\
  SW And &8.19   & +0.09    &7.69   & +0.20&8.60& +0.23&--0.29 \\
  AN Ser &8.37   & +0.00    &7.92   & +0.12&8.71& +0.03& +0.02 \\
  RR Lyr &6.66   &--0.56    &$<$7.50& $<$+0.89&8.30&--0.19&--1.17 \\
  UY Boo &$<$6.10& $<$+0.44 &$<$7.60& $<$+2.55&$<$8.30& $<$+2.37&--2.73 \\
  SU Dra &$<$6.20& $<$--0.25&$<$7.20& $<$+0.36&$<$8.50& $<$+1.78&--1.94 \\
  VY Ser &$<$6.20& $<$--0.53&$<$7.20& $<$+1.08&$<$8.40& $<$+1.40&--1.66 \\
  OX Her &$<$6.50& $<$--0.39&$<$7.20& $<$+0.92&$<$8.50& $<$+1.34&--1.50 \\
 V486 Her&$<$6.90& $<$--0.60&$<$8.00& $<$+1.11&$<$8.30& $<$+0.56&--0.89 \\
  XZ Cet &$<$6.10& $<$--0.51&$<$7.10& $<$+1.10&$<$8.50& $<$+1.62&--1.78 \\
  KP Cyg &9.09& +0.52&8.86&+0.90&8.77&--0.07& +0.18 \\
  UY CrB &8.64& +0.65&8.64&+1.26&8.85& +0.59&--0.40 \\
\hline
   Sun   &8.39 &     &7.78&     &8.66&      &  0.00 \\
\hline
\end{tabular}
\\
 For the Sun we adopted the CNO abundances as recommended by Asplund et al. (2004), Asplund et al. (2005ab).

For UY Boo, SU Dra, VY Ser, OX Her, V486 Her and XZ Cet only upper
limit of CNO abundances is given. For the nitrogen abundance in RR~Lyr
also the upper limit is indicated.
\end{table}

\section{Discussion}

It is well known that in evolving metal-poor stars the first dredge-up
depletes C in the stellar atmosphere, and that the additional mixing
that occurs at the red giant clump further depletes it. Normal or excess
carbon indicates that additional carbon has been injected into the atmosphere
from deeper layers, presumably from the helium flash without being reprocessed
to N by proton capture on the way up. Two stars, KP Cyg and UY CrB, where the carbon
excess in units of [C/Fe] was found to be 0.52 and 0.65, have rather long periods
and [Fe/H] values that are relatively high.

Almost all stars in which it was possible to determine the N abundance show an
excess of that element. Since there is no significant carbon deficiency, the excess
of N must be left over from the first and red giant clump mixing events. Three
stars with [Fe/H] near zero show no N excess.

Only for six stars were we able to measure the oxygen abundance. Except in UY CrB,
there is no indication of an oxygen excess having been generated
during the helium flash by the $^{12}$C $(\alpha$,$\gamma$)$^{16}$O.

Some comments on individual stars are worthwhile.  KP Cyg appears to be one of
the most metal-rich RR Lyrae stars known.  Its carbon abundance is 0.7 dex greater
than solar but its oxygen abundance is not enhanced.
Since its [Fe/H] value is -0.4 UY CrB probably started with logN(C) = 8.0.
It now shows logN(C)=8.6. Its logN(O) must have started near 8.26 and now has
reached 8.85 indicating additional processing of $^{12}$C to $^{16}$O by alpha-capture.
For both stars nitrogen is substantially enhanced as might be expected as
carbon-rich material is subject to proton capture on it way to the surface.
For the metal-poor star, RR Lyr, with [Fe/H] = -1.2 the original carbon abundance
was probably about logN(C) = 7.2 which is likely to have been depleted to 6.7-7.0
at the RG tip and is 7.4 at present. Its present logN(N)=7.4 which must also
be an enhancement from a presumed initial logN(N) = 6.6.

To compare the RR Lyrae stars with less evolved objects whose origin is likely
to be similar we show abundances in various RGB stars in Table 3. For this we have
used data for stars in M71, M4 and 47 Tuc all of which have metallicities
similar to those in the C-rich RR Lyrae stars.

\begin{table}
\scriptsize
\caption[]{CNO abundances in globular clusters}
\begin{tabular}{llllllll}
\hline
Cluster & Type of star&[Fe/H]&$\log$N(C)&$\log$N(N)&$\log$N(O)& $^{12}$C/$^{13}$C&references\\
\hline
M 71    &CNweak       &--0.79& 7.07 & 7.71 & 8.24 & 8.3    &    a   \\
M 71    &CNstrong     &--0.78& 7.11 & 8.07 & 8.08 & 6.0    &    a   \\
M 71    &AGB          &--0.89& 6.76 & 8.04 & 8.14 & 5      &    a   \\
47 Tuc  &GB           &--0.77& 7.48:& 8.29:& 8.35 & 4--12  &    b,d \\
M 4     &GB           &--1.34& 6.97:& 7.83:& 7.87:& 4--8   &    b   \\
M 4     &AGB          &--1.18& 6.64 & 7.35 & 7.73 & 4.5    &    c   \\
\hline
\end{tabular}
\\
References: a - Briley et al. (1997), b - Brown \& Wallerstein (1990), c - Ivans, I.I.,et al. (1999), d- Alves-Brito, et al. (2005).
\end{table}

It appears that some RR Lyrae stars show a significant enhancement in carbon
as compared to red giants of comparable metallicity. Apparently, an increased
carbon abundance at the evolutionary stage after the red giant phase is caused by the deep convective mixing.
Calculations by Fujimoto et al.(2000) and by Schlattl et al.(2002) for
the He-flash in extremely metal-poor stars show that thermally driven
convection in the core can reach the H-burning shell. When H mixes with
the freshly synthesized carbon rapid proton capture can release further
energy to drive the convection to the outer layers of the star.

We are currently expanding our high resolution survey of RR Lyrae stars
to establish any correlation of carbon excesses with period, metallicity
or other parameters.

\begin{acknowledgements}

SMA and VVK would like to express their gratitude for the Kenilworth Fund
of the New York Community Trust for the financial support of this study.
The individual financial support from Kenilworth Fund was made possible through
CRDF.

We thank Marta Mottini and Wenjin Huang for reading the manuscript and making some good suggestions.


The search of the information about the program stars was made using SIMBAD.

\end{acknowledgements}


\begin{thebibliography}{}

  \bibitem[2005]{}
  Alves-Brito, A. et al. 2005 A\&A 435, 657

  \bibitem[1990]{}      
  Anders E., Grevesse N., 1990 Geochim. Cosmochim. Acta 53,197 

  \bibitem[2004]{}
  Asplund M., Grevesse N., Sauval A.J., Allende Prieto C., Kiselman D., 2004,
  A\&A 417, 751

  \bibitem[2005a]{}
  Asplund M., Grevesse N., Sauval A.J., Allende Prieto C., Blomme R., 2005,
  A\&A 431, 693

  \bibitem[2005b]{}
  Asplund M., Grevesse N., Sauval A.J., 2005, ASP Conference Series 336, 25

   \bibitem[1997]{}
   Briley M.M., Smith V.V., Kink J., Lambert D.L., 1997, AJ 113, 306

   \bibitem[1992]{}
   Brown J.A., Wallerstein G. 1992, AJ 100,1561
 
   \bibitem[1995]{}
   Clementini G., Carretta E., Gratton R., Merighi R., Mould J.R., McCarthy J.K.,
   1995, AJ 110, 2319
 
   \bibitem[2000]{}
   Fujimoto M.Y., Ikeda Y., Iben, I.Jr., 2000, ApJ 529, 25

   \bibitem[1992]{}
   Galazutdinov G.A., 1992, Preprint SAO RAS, 92

   \bibitem[1999]{}
   Ivans, I.I. et al. 1999 AJ 118, 1273

   \bibitem[2006]{}
   Dearborn D.S.P., Lattanzio J.C.,  Eggleton P.P., 2006 ApJ 639, 4

   \bibitem[1999]{}
   Kovtyukh V.V.,  Andrievsky S.M., 1999, A\&A 351, 597

  \bibitem[1996]{}
   Kurucz R. 1996, In Model Atmospheres and Spectrum Synthesis,
   ASPC 108, 270

   \bibitem[2008]{}
   Mocak M., Mueller E., Weiss A., Kifonides K., 2008, A\&A, in press

   \bibitem[2002]{}
   Schlattl H., Salaris M., Cassisi S., Weiss A.,  2002 A\&A,
   395, 77


   \bibitem[1985]{}
   Teays T.J., Simon N.R., 1985, ApJ 290, 683

\end{thebibliography}
\end{document}